\documentclass[12pt,russian,showpacs,preprintnumbers]{revtex4}
\input{amssymb.sty}
\usepackage[cp866]{inputenc}
\usepackage{babel}
\usepackage{epsfig}
\usepackage{amsmath}

\newcommand{\Sc}{Schr\"odinger }

\begin{document}

\title{Dynamics of Josephson-junction qubits with exactly solvable
time-dependent bias pulses}
\author{V.V. Shamshutdinova$^{(1)}$, A.S.~Kiyko$^{(2)}$, S.N.~Shevchenko$^{(2)}$,
B.F.~Samsonov$^{(1)}$, A.N.~Omelyanchouk$^{(2)}$}
\affiliation{$^{(1)}$ Tomsk State University, 36 Lenin Avenue, 634050 Tomsk, Russia\\
$^{(2)}$ B. Verkin Institute for Low Temperature Physics and
Engineering, 47 Lenin Ave., 61103, Kharkov, Ukraine}

\begin{abstract}

The quantum dynamics of a two-state system (qubit) can be governed
by means of external control parameters present in time-dependent
bias pulses of special forms. We consider the class of biases for
which the time evolution equation without a dissipation can be
solved exactly. Concentrating for definiteness on the flux qubit
we calculate the probability of the definite direction of the
current in the loop and its time-averaged values as functions of
the qubit's control parameters both analytically and solving
numerically the equation of motion for the density matrix in the
presence of relaxation and decoherence. It is shown that there
exist such time-dependent biases that the definite current
direction probability with no dissipation taken into account
becomes a monotonously growing function of time tending to a value
which may exceed 1/2. We also calculate the probability to find
the system in the excited state and show the possibility of the
inverse population in a properly driven two-state system provided
the relaxation and dephasing rates are small enough.

\end{abstract}
\pacs{03.67.Lx, 03.75.Lm, 85.25.Am}

\maketitle

\section{Introduction}

In the past few years Josephson junctions-based devices have been
widely studied both theoretically and experimentally as possible
candidates for the implementation of a quantum computer
\cite{FromEPJB34(269)1}-\cite{FromEPJB34(269)9}. In fact, under
appropriate values of external bias pulses, they behave as
two-state systems which can be used as a model for quantum
bits~(qubits). Several systems like ion traps and NMR systems
\cite{FromEPJB34(269)10}, \cite{FromEPJB34(269)11} have been
suggested for physical realizations of the qubit but Josephson
devices being scalable up to large numbers of qubits as
nanocomponents embedded into an electronic circuit exhibit the
main technological advantage. Moreover, it is possible to prepare
these devices in a prefixed initial state or in a superposition of
states and to control their dynamics by an external voltage and
magnetic flux \cite{FromEPJB34(269)12}.

One common approach to control the qubit dynamics is to drive the
two-state system, i.e. a particle in a double well potential, with
an oscillating field. As a result an interesting phenomenon may
take place. Instead of oscillating between the wells the particle
may become localized in one of them. Such an unusual behavior of
the quantum particle is known in the literature as coherent
destruction of tunneling in double-well potential studies
\cite{Grossmann67}-\cite{Gomes}, dynamic localization in transport
analysis \cite{Raghavan},~\cite{Dunlap} and population trapping in
laser-atom physics \cite{Agarwal}. It is worth stressing that up
to now the phenomenon was essentially related with an oscillating
character of the driving external field.

In the present article we show that the oscillating character of
the field is not compulsory for appearance of the phenomenon.
We present a class of non-periodical time-dependent bias pulses
leading to a similar behavior of the qubit.

A qubit in the two-state approximation can be described (see e.g.
\cite{FromEPJB34(269)12},~\cite{Reviews}) by the \Sc equation
\begin{equation*}
i\frac{\partial\Psi}{\partial t}=H\Psi
\end{equation*}%
with the Hamiltonian
\begin{equation}\label{eq}
H=\Delta\sigma_{x}+\varepsilon\left(t\right)\sigma_{z}\,,\quad
\hbar=1
\end{equation}%
written down in the basis of \textquotedblleft
physical\textquotedblright\ states $\{\left\vert 0\right\rangle
,\left\vert 1\right\rangle\}$, which are the eigenstates of the
Pauli matrix $\sigma_{z}$ ($\sigma_{z}\left\vert 0\right\rangle
=\left\vert 0\right\rangle$, $\sigma_{z}\left\vert 1\right\rangle
=-\left\vert 1\right\rangle$) and $\Psi =\left(\psi_{1},\psi
_{2}\right)^{T}$. In the case of a charge
qubit~\cite{FromEPJB34(269)1}, these states correspond to a
definite number of Cooper pairs on the island (Cooper-pair box).
For a flux qubit \cite{fluxQbit}, they correspond to a definite
direction of the current circulating in the ring. In this paper we
assume that the tunnelling amplitude $\Delta $ is constant and the
bias $\varepsilon$ is a time-dependent function,
$\varepsilon=\varepsilon(t)$. Bias~$\varepsilon(t)$ is governed by
gate voltage $V_{g}(t)$ of the gate electrode close to the
Cooper-pair box in the case of a charge qubit and by magnetic
flux~$\Phi_{x}(t)$ piercing the qubit's loop in the case of a flux
qubit.

For definiteness we will consider here flux qubits. Then
\textquotedblleft physical\textquotedblright\ states~$\{\left\vert
0\right\rangle,\left\vert 1\right\rangle\}$ are the states with
the definite (clockwise or counter-clockwise) direction of the
current in the loop. The above mentioned time-dependent biases
are, in fact, potentials for which Schr\"odinger equation
\eqref{eq} can be solved exactly
\cite{Shamshutdinova}-\cite{Bagrov}. Thus, the probability
calculated using these exact solutions is, for example,
probability $P^{\uparrow}$  of clockwise current direction. We
demonstrate that for some special non-periodical forms of
potentials the probability of the clockwise current direction at
the moment $t$, if at~$t=0$ it was counter-clockwise, becomes a
monotonously growing function of time tending to~3/4. Of course
this is a strictly fixed excitation regime but we also study the
behavior of the probability under small deviations from this
specific regime. It should be noted that when parameters of the
model are close enough to their specific values, the probability
oscillates but its minimal value may exceed 1/2. It is established
that such an unusual behavior of the probability is possible even
in the presence of a dissipation. We study not only the time
evolution of the probability, but also its time-averaged value.
Moreover, we also calculate the probability of finding the system
in the excited state and show the possibility of the inverse
population in the two-state system even in the presence of a
dissipation. Our main result is that using a properly chosen
non-periodical time-dependent potentials one can ``freeze''  the
qubit state i.e. localize only one of the two possible qubit
states for a long time interval. We note that the probability of
the definite current flow (or equivalently, the definite magnetic
moment of the qubit) is related directly to experimentally
measurable values such as the phase shift of the resonant circuit,
weakly coupled to the qubit (as discussed, e.g., in Ref.
\cite{KOSh06}). Moreover, this \textquotedblleft
physical\textquotedblright\ basis is usually used in quantum
computations. Thus, we hope that our results offer new
opportunities for controlling the qubit behavior. Additional
discussion about controlling the level population can be found in
Refs. \cite{Siewert}-\cite{Liu}.

\section{Exactly solvable bias pulses}

In order to construct an exact solution of a differential equation
the intertwining operator technique sometimes may be useful. The
idea of the method dates from Darboux papers~\cite{Darb} and is
widely used in the soliton theory \cite{Matv}. Its quantum
mechanical application (see e.g.~\cite{BS}) is related to the fact
the one-dimensional stationary \Sc equation is an ordinary second
order differential equation defined by the operator of the
potential energy. The method is based on the possibility to find
an operator (intertwining operator) that relates solutions of two
\Sc equations with different potentials. Thus, if one knows
solutions of the \Sc equation with a given potential and an
intertwining operator is available, there exists a possibility to
construct solutions of the same equation with another potential,
which cannot be completely arbitrary but is an internal
characteristic of the method. There exists also a
matrix-differential formulation of the method \cite{Matv} which
was adapted to quantum mechanical problems in Ref. \cite{Nieto}.

In \cite{Shamshutdinova} it was shown how to construct
differential-matrix intertwining operators for the system of two
differential equations of type \eqref{eq}. For that authors
\cite{Shamshutdinova} first reduce system~\eqref{eq} to the
one-dimensional stationary Dirac equation with an effective
non-Hermitian Hamiltonian where the time plays the role of a space
variable and then apply the known procedure developed in
\cite{Nieto}. Starting from the simplest case corresponding to
$\varepsilon=\varepsilon_0=\mbox{const}$, a new kind of nontrivial
potentials (biases) for which Schr\"odinger equation \eqref{eq}
can be solved exactly were found. Here we apply results obtained
in~\cite{Shamshutdinova}-\cite{Bagrov} to describe the time
evolution of the qubit, time dependence of the qubit localization
probability and calculate its time-averaged value.

\begin{figure}[p]
\begin{minipage}{15.5cm}
\includegraphics[width=7cm]{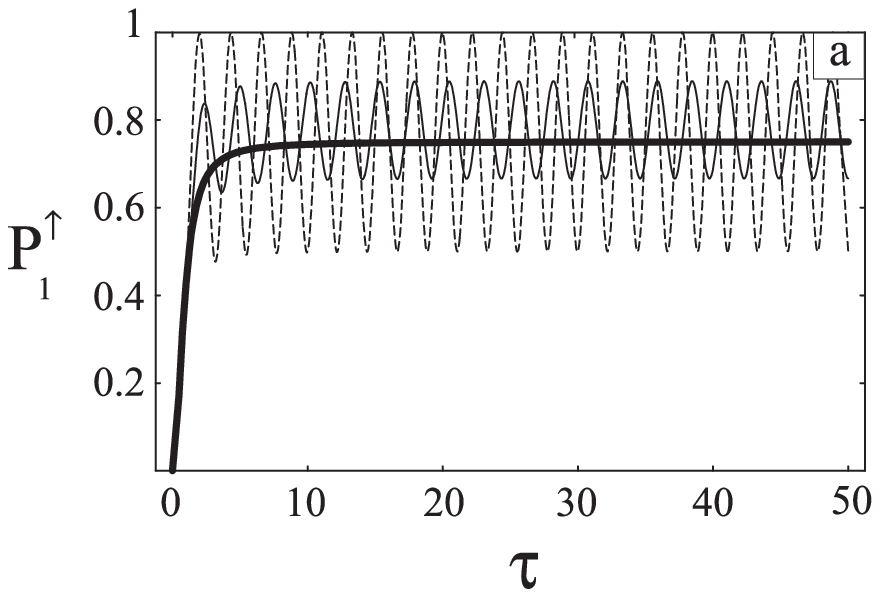}
\hspace{1cm}
\includegraphics[width=7cm]{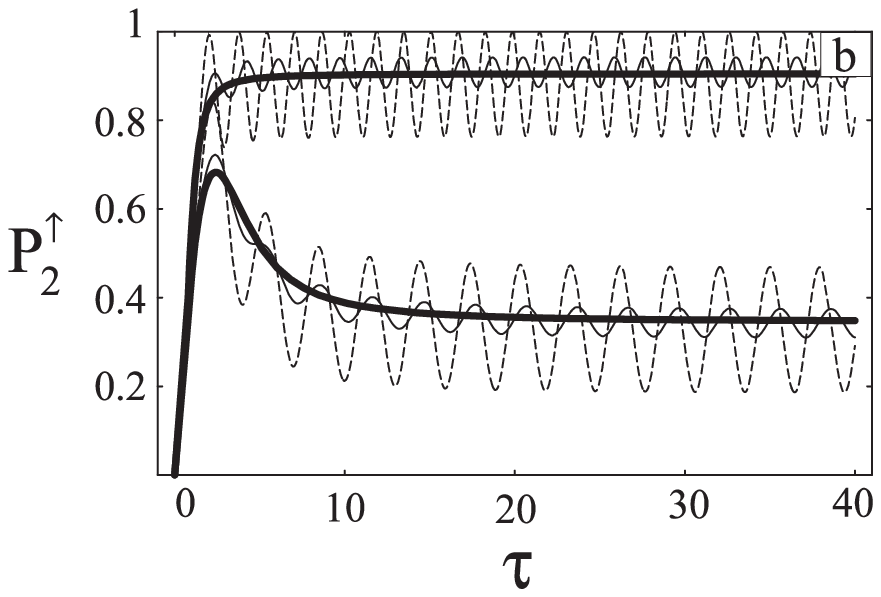}
\caption{Time dependence of clockwise current direction
probabilities. (a) $ P_1^\uparrow$ probability evolution at
$\xi=\sqrt{\frac{1}{2}}$ (thin line), $\xi=1$ (dotted line) and
$\xi=\sqrt{\frac{1}{3}}$ (thick line). (b) $ P_{2}^{\uparrow }$
probability evolution at $\xi=\sqrt{2+\frac{2}{\sqrt{5}}}$,
$\xi=\sqrt{1.2+\frac{2}{\sqrt{5}}}$ and
$\xi=\sqrt{1+\frac{2}{\sqrt{5}}}$ on the upper plot,
$\xi=\sqrt{1.05-\frac{2}{\sqrt{5}}}$,
$\xi=\sqrt{1.01-\frac{2}{\sqrt{5}}}$ and $
\xi=\sqrt{1-\frac{2}{\sqrt{5}}}$ on the lower plot (dotted, thin
and thick lines respectively).}\label{fig1a}\label{fig1b}
\end{minipage}
\end{figure}

Consider first the case when bias $\varepsilon=\varepsilon(t)$
changes in the following way:
\begin{equation}
\varepsilon_{1}\left(t\right) =\varepsilon _{0}-\frac{4\varepsilon _{0}}{%
1+4\varepsilon _{0}^{2}t^{2}}\;.  \label{e1}
\end{equation}%
In Ref. \cite{Shamshutdinova} a detailed analysis of solutions to equation %
\eqref{eq} in this case is given. Therefore imposing the initial
conditions $|\psi _{1}\left(0\right)|^2=1$ and
$|\psi_{2}\left(0\right)|^2=0$ one finds probability
$P^{\uparrow}\left(t\right)=|\psi_{2}\left(t\right)|^2$ of, for
example, the clockwise current direction at the moment $t$ if
at~$t=0$ it was counter-clockwise. For $\tau=\Delta t$ and
$\xi=\frac{\varepsilon_{0}}{\Delta}$ it reads
\begin{multline}
P_{1}^{\uparrow}\left(\tau\right)=\frac{1}{\Theta^{6}\left(
1+4\xi^{2}\tau^{2}\right)}\times\\
\left[16\xi^{4}\Theta^{2}\tau^{2}\cos^{2}\Theta\tau+4\xi
^{2}\Theta \tau \left( 1-3\xi ^{2}\right) \sin 2\Theta \tau
+\left(4\xi^{2}\Theta^{4}\tau^{2}+\left(1-3\xi ^{2}\right)
^{2}\right) \sin ^{2}\Theta \tau \right] ,
\end{multline}
\begin{equation*}
\Theta=\sqrt{1+\xi^{2}}\;.
\end{equation*}
It is clearly seen from here that $P_{1}^{\uparrow}(\tau)$ is an
oscillating function provided $\xi ^{2}\neq \frac{1}{3}$.
For~$\xi^{2}=\frac{1}{3}$ the probability becomes equal
\begin{equation}
P_{1}^{\uparrow}\left(\tau \right)=\frac{\tau ^{2}}{1+\frac{4}{3}\tau ^{2}}%
\;,  \label{P_mon}
\end{equation}
which is a function monotonically growing from zero at the initial
time moment till the value~$3/4$ at $\tau\gg 1$ or
$t\gg\Delta^{-1}$ (see the thick line in Fig. \ref{fig1a}a). It is
important to note that at~$\xi^{2}$ close enough to $\frac{1}{3}$
the value of probability $P_{1}^{\uparrow }$ exceeds $1/2$ very
quickly (see the thin and dotted lines in Fig. \ref{fig1a}a).

\begin{figure}[p]
\begin{minipage}{15.5cm}
  \includegraphics[width=7cm]{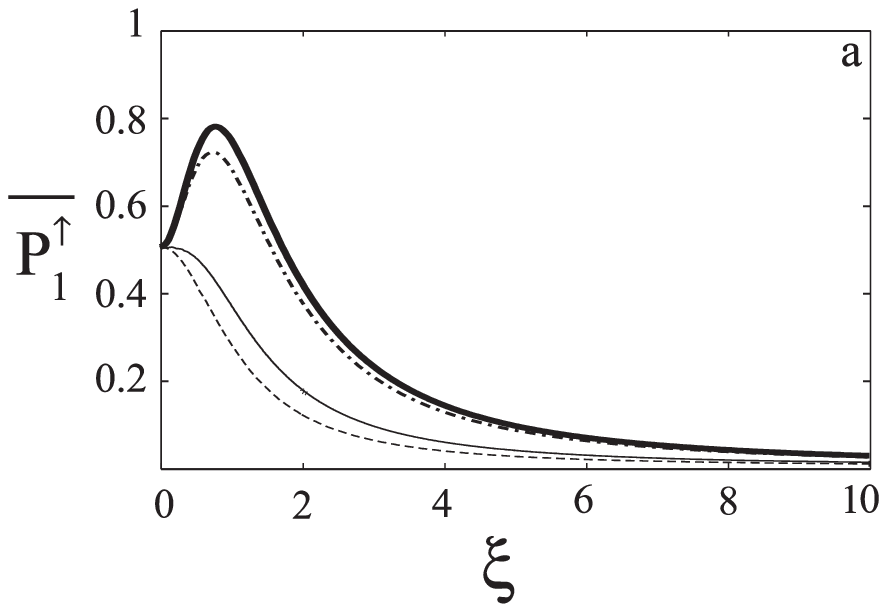}
\hspace{1cm}
\includegraphics[width=7cm]{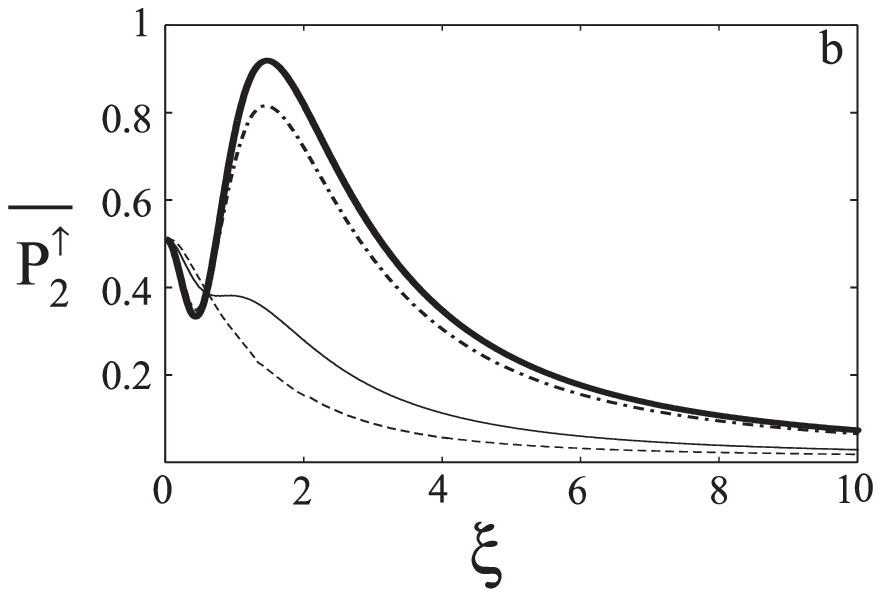}
  \caption{$\xi$-dependence of time-averaged
probabilities.}\label{fig5_a}\label{fig6_a}
 \end{minipage}
\end{figure}

For the time-averaged probability one gets
\begin{equation}
\overline{P_{1}^{\uparrow}}=\frac{1+5\xi^{2}}{2\left( 1+\xi^{2}\right)^{2}}%
\;.  \label{P1aver1}
\end{equation}
It follows from Eq. \eqref{P1aver1} that at $\xi^2=\frac{3}{5}$
the averaged probability exhibits a kind of the resonance
behavior, i.e. it peaks to its maximal value
$\overline{P_{1}^{\uparrow }}\approx 0.78$, (see the thick line in
Fig.~\ref{fig5_a}) although for any given $\xi$
$P_{1}^{\uparrow}\left(\tau\right) $ is a function asymptotically
oscillating with frequency $4\sqrt{\frac{2}{5}}$
(see~Fig.~\ref{fig1a}).

More exactly solvable potentials may be obtained with the help of
chains of the above simple transformations. Ref. \cite{PolPSUSY}
contains a detailed analysis of the properties of such chains. The
authors show how to generate a large family of new exactly
solvable biases for equation~\eqref{eq}. For a two-fold
transformation leading to the bias of the form
\begin{equation} \label{e2}
\varepsilon _{2}\left( t\right) =\frac{\varepsilon _{0}\left(
45-180\varepsilon _{0}^{2}t^{2}-144\varepsilon
_{0}^{4}t^{4}+64\varepsilon _{0}^{6}t^{6}\right)
}{9+108\varepsilon _{0}^{2}t^{2}+48\varepsilon
_{0}^{4}t^{4}+64\varepsilon _{0}^{6}t^{6}}
\end{equation}%
the clockwise current direction probability is given by
\begin{eqnarray}
P_{2}^{\uparrow }\left( \tau \right) & = &\frac{1}{\Theta ^{10}\left(
9+108\xi ^{2}\tau ^{2}+48\xi ^{4}\tau ^{4}+64\xi ^{6}\tau
^{6}\right) }\notag
 \\ \notag
& \times & \biggl(16\xi ^{4}\Theta ^{2}\tau ^{2}\left[ 16\xi ^{4}\Theta
^{4}\tau ^{4}+24\xi ^{2}\left( 3-14\xi ^{2}+7\xi ^{4}\right) \tau
^{2}+9\left( 9-6\xi ^{2}+\xi ^{4}\right) \right]\\
& + & Q_{1}\left[ Q_{2}\sin ^{2}\Theta \tau +Q_{3}\sin 2\Theta \tau
\right] \biggr)\label{P2e2}
\end{eqnarray}
where
\begin{equation*}
\begin{array}{l}
Q_{1}=1-10\xi ^{2}+5\xi ^{4}\,, \\
Q_{2}=64\xi ^{6}\Theta ^{4}\tau ^{6}+48\xi ^{4}\left( 1-18\xi
^{2}-19\xi ^{4}\right) \tau ^{4}+36\xi ^{2}\left( 3-2\xi
^{2}+11\xi ^{4}\right) \tau ^{2}+9Q_{1}\,,\\
Q_{3}=12\xi ^{2}\Theta \tau \left( \Theta ^{2}\left( 16\xi
^{4}\tau ^{4}+7\right) +2\left( 4\xi ^{2}\tau ^{2}+1\right) \left(
1-5\xi ^{2}\right) \right) .
\end{array}
\end{equation*}
The last term in Eq. (\ref{P2e2}) describes the oscillations with
frequency $2\Theta$ and, hence, under the condition $Q_{1}=0$ the
oscillations in the time-dependence of the probability disappear
and it acquires a monotonous character. Therefore, for
$\varepsilon $ as given in (\ref{e2}), unlike (\ref{e1}), we can
indicate two possibilities for $Q_{1}=0$. So, the probability of
the clockwise current direction turns from an oscillating to a
monotonous function of time both at $\xi =\sqrt{1-2/\sqrt{5}}$ and
$\xi =\sqrt{1+2/\sqrt{5}}$. This behavior is illustrated in
Fig.~\ref{fig1b}b (thick lines) where we also show an oscillating
character of the probability for the parameter $\xi$ close to the
above critical values (thin and dotted lines).

Time-averaged probability (\ref{P2e2}) for
$\varepsilon\left(t\right) $ of the form (\ref{e2}) is given by
\begin{equation}
\overline{P_{2}^{\uparrow }}=\frac{1-2\xi ^{2}+13\xi ^{4}}{2\left(
1+\xi ^{2}\right) ^{3}}\,.
\end{equation}%
The $\xi$-dependence of $\overline{P_{2}^{\uparrow }}$  is
demonstrated in Fig. \ref{fig5_a}b by the thick line. It has a
maximum $\overline{P_{2}^{\uparrow }}\approx 0.91$, at $\xi
\approx 1.46$.

Let us consider a more complicated case \cite{Shamshutdinova},
\cite{Bagrov} when the bias contains three free parameters
\begin{equation}
\varepsilon _{3}\left( t\right) =\varepsilon _{0}+\frac{2\omega
^{2}}{b\cos \left( 2\omega t+\varphi \right) -\varepsilon
_{0}}\;,\quad b^{2}=\varepsilon _{0}^{2}-\omega ^{2}>0\,.
\label{e3}
\end{equation}%
Here $\varphi $ is arbitrary but $\varepsilon _{0}$ and $\omega $
should satisfy the inequality given in Eq. \eqref{e3}. We note
that in this case $\varepsilon =\varepsilon _{3}(t)$ is a
periodical function with an amplitude related with frequency.
Expression (\ref{e3}) presents a generalization of formula
(\ref{e1}). Indeed, putting
$\varphi=\arctan\frac{\omega}{2\varepsilon_{0}}-\frac{1}{%
2}\arctan\frac{\omega}{b}$ in Eq.~(\ref{e3}), in the limit $\omega
\rightarrow 0$ one recovers for $\varepsilon $ result (\ref{e1}).

\begin{figure}[p]
\begin{minipage}{15.5cm}
\includegraphics[width=7cm]{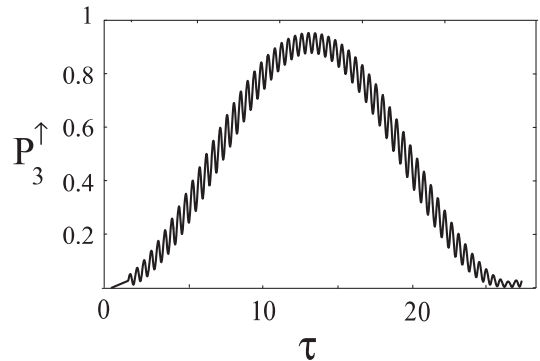}
\caption{Time dependence of clockwise current direction
probabilities. $P_{3}^{\uparrow}$ probability evolution at $\theta
=6.88$, $\Theta=7$, $\xi=\sqrt{48}$, $\varphi=0$.}\label{fig3}
\end{minipage}
\end{figure}

The analytic expression for $P_{3}^{\uparrow }(\tau)$ is rather
involved and we will restrict ourselves to a graphical
illustration of the clockwise current direction probability at
$\Theta\approx\frac{\omega}{\Delta}$, (see Fig.~\ref{fig3}). More
graphical illustrations can be found in \cite{Shamshutdinova}.

\section{Influence of a dissipation on probabilities}

A quantum system described by a wave function which is a solution
of the \Sc equation with Hamiltonian \eqref{eq} interacts only
with an external field described by function~$\varepsilon(t)$. But
in real experiments there is also an interaction of a Josephson
junction device with an external reservoir which makes impossible
describing the system in terms of a wave function since its state
is not a pure quantum state anymore and should be described withe
the help of a density matrix (operator in general see e. g
\cite{Breuer}) formalism. Such kind of systems are particularly
important in the context of quantum information processing where
environment-induced decoherence is viewed as a fundamental
obstacle for constructing a quantum information processors (e.g.,
\cite{Lidar}).

In this section we study the behavior of the qubit with bias
$\varepsilon (t)$ as given in Eqs.~\eqref{e1},~\eqref{e2} and
\eqref{e3} taking into account the dephasing and the relaxation
processes. We also study a possibility of the inverse population
in the two-level system and investigate how it is influenced by a
decoherence. It is worth noticing that in contrast to Ref.
\cite{Siewert}, where the authors investigate a three-level
system, we study a possibility of the inverse population for the
two level system itself thus showing the possibility of building a
two-level based laser working at a low temperature.

To study the dynamic behaviour of the qubit, we use the master
equation for the density matrix. For the density matrix of the
form
\begin{equation*}
\widehat{\rho }=\frac{1}{2}\left[
\begin{array}{cc}
1+Z & X-iY \\
X+iY & 1-Z%
\end{array}%
\right]
\end{equation*}%
we solve the equation of motion
\begin{equation*}
i\frac{\partial \widehat{\rho }}{\partial
t}=[\widehat{H}\widehat{\rho }]
\end{equation*}%
to obtain the probability $P^{\uparrow}=[1-Z(t)]/2$. The effect of
the relaxation processes on the system due to a weak coupling to
the environment can be phenomenologically described by two
parameters, the dephasing ($\Gamma_{\varphi }$) and relaxation
($\Gamma_{relax}$) rates (see e.g. Ref. \cite{ShKOK}), which we
introduce phenomenologically thus obtaining the following system
of equations for $X(t)$, $Y(t)$ and $Z(t)$:
\begin{align}
\frac{dX}{dt}& =-2\varepsilon (t)Y-\Gamma _{\varphi }X,  \notag \\
\frac{dY}{dt}& =-2\Delta Z+2\varepsilon (t)X-\Gamma _{\varphi }Y,  \notag \\
\frac{dZ}{dt}& =2\Delta Y-\Gamma _{relax}\left( Z-Z(0)\right) .
\notag
\end{align}%
In order to verify the possibility of the inverse population in
the two-level system we also calculate the probability $P^{+}$
\cite{ShKOK} of finding the system on the upper level (excited
state).

\begin{figure}[p]
\begin{minipage}{15.5cm}
  \includegraphics[width=7cm]{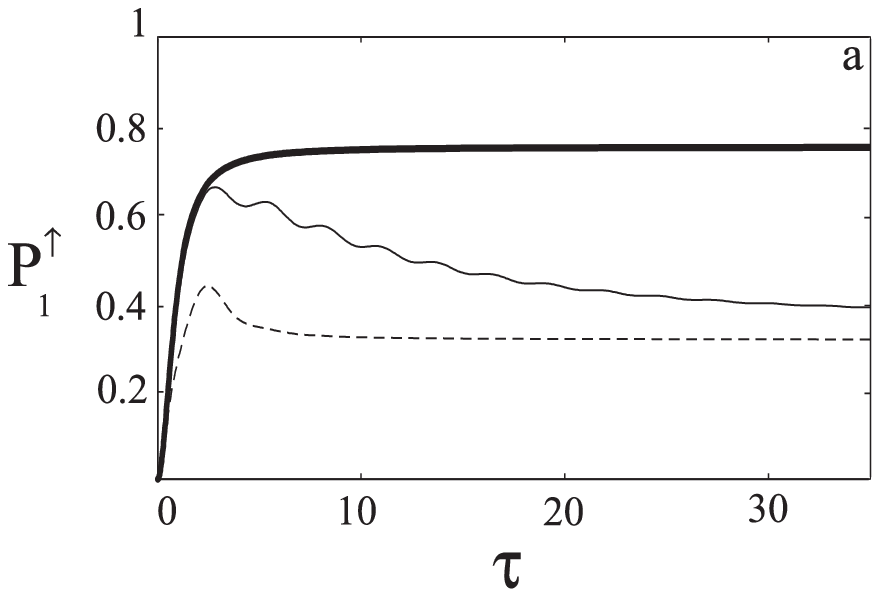}
\hspace{1cm}
 \includegraphics[width=7cm]{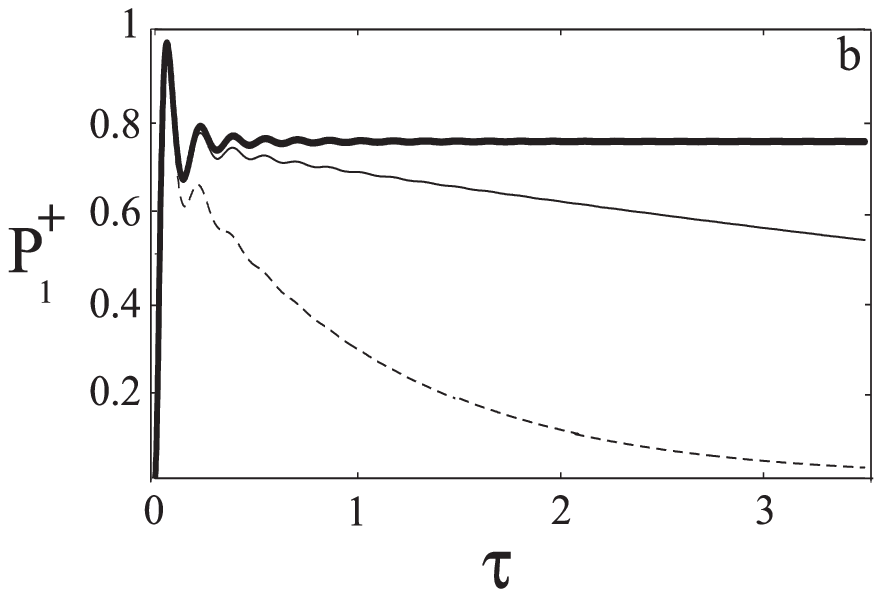}
  \caption{The influence of relaxation on
the probabilities. (a) The clockwise current direction probability
$P_{1}^{\uparrow}$ at $\protect\xi =1/\protect\sqrt{3\text{.}}$
(b) The upper level occupation probability $P_1^{+}$ at
$\protect\xi
=1/\protect\sqrt{3\text{.}}$}\label{fig1_a}\label{fig2_a}
 \end{minipage}
\end{figure}

\begin{figure}[p]
\begin{minipage}{15.5cm}
  \includegraphics[width=7cm]{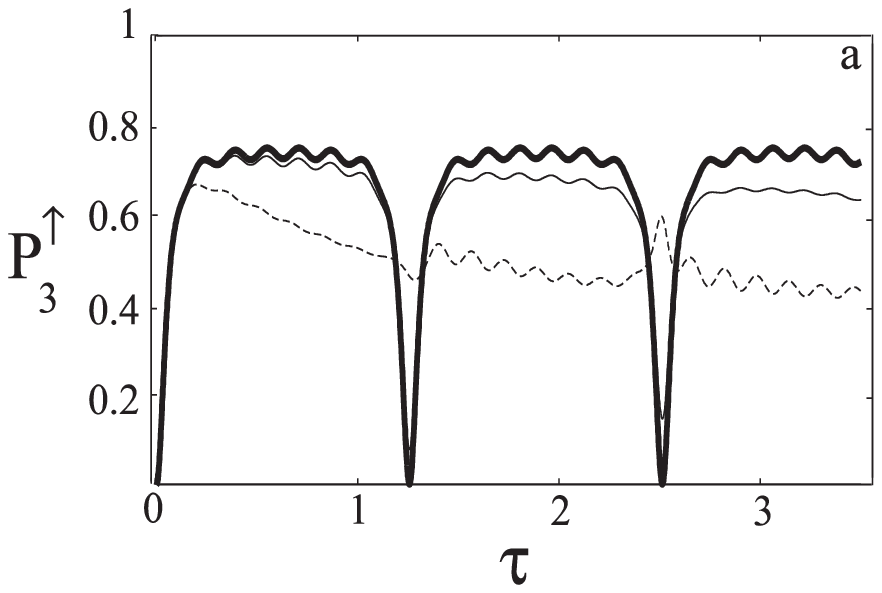}
\hspace{1cm}
 \includegraphics[width=7cm]{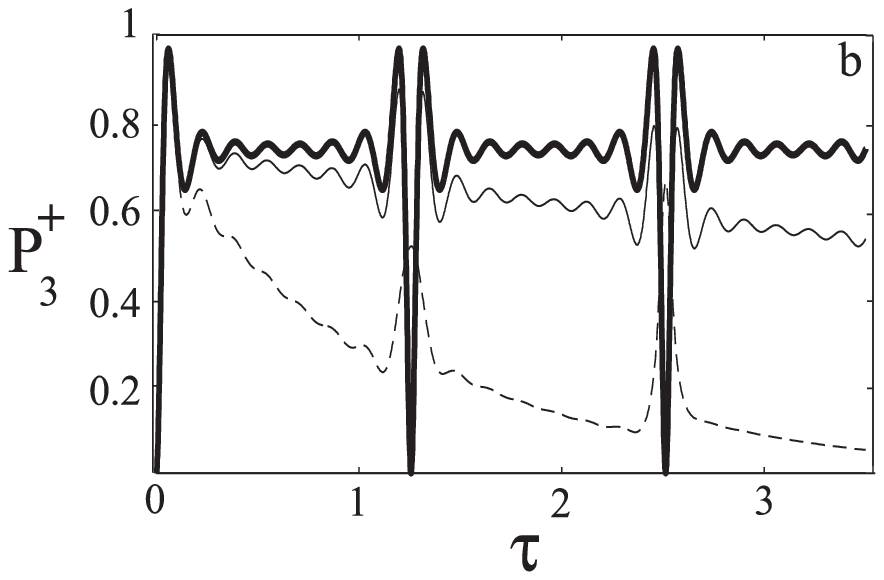}
  \caption {The influence of relaxation on
the probabilities. (a) The clockwise current direction probability
$P_{3}^{\uparrow }$ at $\protect\xi =\frac{1}{\protect\sqrt{3}}$,
$b=\frac{\protect\sqrt{15}}{2}$, $\protect\varphi =0$. (b) The
upper level occupation probability $P_3^{+}$ at $\protect\xi
=\frac{1}{\protect\sqrt{3}}$, $b=\frac{\protect\sqrt{15}}{2}$,
$\protect\varphi =0$.}\label{fig3_a}\label{fig4_a}
\end{minipage}
\end{figure}

To see the influence of the relaxation on probabilities
$P_{1}^{\uparrow }$ and $P_{1}^{+}$ exhibiting a monotonous time
dependence for $\varepsilon(t)$ in form \eqref{e1} we choose $\xi
=1/\sqrt{3}$ and plot  them in Fig. \ref{fig1_a}. We do not show
the behavior of  $P_{2}^{\uparrow}$ and $P_{2}^{+}$ corresponding
to $\varepsilon(t)$ as given in \eqref{e2} since it is similar to
that displayed in Fig. \ref{fig1_a}. Fig. \ref{fig3_a} illustrates
the evolution of probabilities $P_{3}^{\uparrow}(t)$ and
$P_{3}^{+}$ when $\varepsilon (t)$ has form \eqref{e3} with $\xi
=\frac{1}{\sqrt{3}}$, $\beta=\frac{\sqrt{15}}{2}$, $\varphi=0$.
Thick lines on these figures correspond to $\Gamma_{\varphi
}=\Gamma _{relax}=0$ when no relaxation is present in the system.
Thin and dotted lines just show the relaxation effect for
$\Gamma_{\varphi }=\Gamma_{relax}=0.01$ and $\Gamma_{\varphi
}=\Gamma_{relax}=0.1$ respectively. All values are in units of
$\Delta $. It is clearly seen from here that the inverse
population is still possible even if the dissipation is present
but during a small time interval $\tau $ only.

Time-averaged probabilities $\overline{P_{1}^{\uparrow}}$ and
$\overline{P_{2}^{\uparrow}}$ are plotted in Fig. \ref{fig5_a}a
and \ref{fig5_a}b as functions of dimensionless parameter $\xi$.
Here the thick lines also correspond to the absence of the
relaxation and the thin and dotted lines illustrate the relaxation
effect for the same values of $\Gamma_{\varphi}=\Gamma_{relax}$ as
in Figs. \ref{fig1_a} and \ref{fig3_a}. The dash-dotted curves are
drown for $\Gamma_{\varphi}/\Delta=\Gamma_{relax}/\Delta =0.001$.
These results show the possibility of the inverse population only
for small enough dissipation (see the dash-dotted lines) and for
parameter $\xi$ close enough to its critical values when the
averaged probabilities pick to their maximal values.

\section{Conclusion}

In conclusion, a two-state system subjected to biases of special
forms, for which the Schr\"{o}dinger equation is exactly solvable,
was considered. For definiteness we studied the time-evolution of
the Josephson-junction qubit, but the similar consideration can be
applied to other realizations of a two-state system. We have
demonstrated that the dynamics of the Josephson-junction qubit
displays in these cases several nontrivial features. Varying the
form of the time dependent bias, one can change qualitatively the
dynamic behavior of the occupation probabilities. In particular,
the amplitude of the oscillating probability, describing the
definite current direction in the loop, can be tuned to zero, thus
turning the probability to a  monotonous function of time. We
calculate the probability of finding the system in the excited
state and show that the inverse population in the two-state system
is possible even in the presence of a dissipation. The observation
of such an occupation probability behavior may be related to
experimentally measurable values, which would make an experimental
verification of the theoretical predictions of the present paper.

V.V.S. acknowledges a partial support from INTAS Fellowship Grant
for Young Scientists Nr~06-1000016-6264. S.N.S and A.N.O.
acknowledge the grant \textquotedblleft Nanosystems,
nanomaterials, and nanotechnology\textquotedblright\ of the
National Academy of Sciences of Ukraine. The work of S.N.S. was
partly supported by grant of President of Ukraine (No. GP/P11/13).
The work of B.F.S. is partially supported by grants
RFBR-06-02-16719 and SS-5103.2006.2.

\end{document}